\documentclass[a4paper,12pt]{article}
\pdfoutput=1

\usepackage[height=23truecm,width=16truecm,top=3.05truecm]{geometry}

\usepackage[utf8]{inputenc}
\usepackage{graphicx}
\usepackage{amsmath}
\usepackage{amssymb}
\usepackage{hyperref}
\hypersetup{colorlinks,bookmarksopen,bookmarksnumbered,citecolor=TokiwaIro,
linkcolor=RuriIro,urlcolor=RuriIro,pdfstartview=FitH,linktocpage}
\usepackage{cite}
\usepackage{braket}
\usepackage{bm}
\usepackage{color}
\usepackage[svgnames,psnames,table]{xcolor}
\usepackage{float}
\usepackage{hhline}
\usepackage{multirow}
\usepackage{indentfirst}
\usepackage{latexsym}
\usepackage{mathtools}
\usepackage{cancel}
\usepackage{slashed}
\usepackage{colortbl}
\usepackage[mathscr]{euscript}
\usepackage{enumerate}

\numberwithin{equation}{section}

\allowdisplaybreaks

\definecolor{RuriIro}{rgb}{0.,0.28,0.60}
\definecolor{TokiwaIro}{rgb}{0.,0.39,0.16}

\newcommand{\nn}{\nonumber}

\newcommand{\bmat}[1]{\begin{bmatrix}#1\end{bmatrix}}

\newcommand{\mc}{\mathcal}
\newcommand{\mr}{\mathrm}

\newcommand{\mbb}{\mathbb}

\newcommand{\del}{\partial}
\newcommand{\ol}{\overline}
\def\tr{\mathop{\mathrm{tr}}\nolimits}
\def\str{\mathop{\mathrm{str}}\nolimits}

\def\Re{\mathop{\mathrm{Re}}}
\def\Im{\mathop{\mathrm{Im}}}

\def\diag{\mathop{\mathrm{diag}}}

\def\Vol{\mathop{\mathrm{Vol}}}

\newcommand{\ls}{\ell_s}
\newcommand{\ap}{\alpha'}

\newcommand{\DBI}{\text{DBI}}

\newcommand{\NDBI}{\text{NDBI}}

\newcommand{\fm}[2]{m^{(#1)}_{#2}}
\newcommand{\fM}[2]{M^{(#1)}_{#2}}
\newcommand{\fI}[2]{I^{(#1)}_{#2}}
\newcommand{\bF}{\hat{G}}

\renewcommand{\bar}{\overline}

\begin{document}

\begin{titlepage}

\begin{flushright}
KUNS-2882\\
EPHOU-21-007
\end{flushright}

\vspace{1cm}

\begin{center}

{\Large \bfseries
4D effective action from non-Abelian DBI action\\
with magnetic flux background
}

\vspace{1cm}

\renewcommand{\thefootnote}{\fnsymbol{footnote}}
{%
\hypersetup{linkcolor=black}
Yoshihiko Abe$^{1}$\footnote[1]{y.abe@gauge.scphys.kyoto-u.ac.jp},
\ 
Tetsutaro Higaki$^{2}$\footnote[2]{thigaki@rk.phys.keio.ac.jp},
\ 
Tatsuo Kobayashi$^{3}$\footnote[3]{kobayashi@particle.sci.hokudai.ac.jp},
\\
Shintaro Takada$^{3}$\footnote[4]{s-takada@particle.sci.hokudai.ac.jp}
\ 
and
\ 
Rei Takahashi$^{2}$
}%
\vspace{5mm}

{\itshape%
$^1${Department of Physics, Kyoto University, Kyoto 606-8502, Japan}\\
$^2${Department of Physics, Keio University, Yokohama 223-8533, Japan}\\
$^3${Department of Physics, Hokkaido University, Sapporo 060-0810, Japan}
}%

\vspace{8mm}

\abstract{
We study a systematic derivation of
four dimensional $\mathcal{N}=1$ supersymmetric effective theory from ten dimensional non-Abelian Dirac-Born-Infeld action compactified on a six dimensional torus with magnetic fluxes on the D-branes. 
We find a new type of matter K\"{a}hler metric while gauge kinetic function and superpotential are consistent with previous studies.
For the ten dimensional action, we use a symmetrized trace prescription and focus on the bosonic part up to $\mathcal{O}(F^4)$.
In the presence of the supersymmetry, four dimensional chiral fermions can be obtained via index theorem.
The new matter K\"{a}hler metric is independent of flavor but depends on the fluxes, 4D dilaton, K\"{a}hler moduli and complex structure moduli, and will be always positive definite if an induced Ramond-Ramond charge of the D-branes on which matters are living are positive.
We read the superpotential from an F-term scalar quartic interaction derived from the ten dimensional action and the contribution of the new matter K\"{a}hler metric to the scalar potential which we derive turns out to be consistent with the supergravity formulation. 
}

\end{center}
\end{titlepage}

\renewcommand{\thefootnote}{\arabic{footnote}}

\setcounter{footnote}{0}

\setcounter{page}{1}

\tableofcontents

\section{Introduction}
\label{sec:intro}

Superstring theory is an attractive candidate for 
a unified theory
consistent with quantum gravity.
The theory can provide us with a theoretical framework to 
describe all the interactions and chiral matters such as quarks and leptons as well as the Higgs field.
The string theory can predict
the existence of extra dimensions
and D-branes.
Dynamics of low energy excitations on D-branes is 
described by gauge theories.
Compactification of string theory on tori is one of simple ways to obtain four-dimensional (4D) effective field theories but these are nonchiral,
while the Standard Model is chiral.
The chiral nature of matter fields is realized by
introducing magnetic fluxes
on the world volume of D-branes in the compact extra dimensions
\cite{Bachas:1995ik,Blumenhagen:2000wh,Angelantonj:2000hi,Blumenhagen:2000ea}.
Even in toroidal compactfications, magnetic fluxes realize 4D chiral theory.
Orbifold compactification with magnetic fluxes is also studied in Refs.~\cite{Abe:2008fi,Abe:2013bca,Abe:2014noa}.
The number of chiral generations is determined by the size of the magnetic flux on compact extra dimensions.\footnote{
The number of chiral generations also depends on twisted boundary conditions, discrete Wilson lines, and 
Scherk-Schwarz
phase in orbifold models.
}
Three-generation models have been classified in Refs.~\cite{Abe:2008sx,Abe:2015yva,Hoshiya:2020hki}.
Moreover, as the zero-mode functions of the Dirac (Laplace) operator
are quasilocalized in compact space 
and Yukawa couplings as well as higher order couplings are written
by overlap integration among their zero mode functions,
hierarchical couplings can be realized \cite{Cremades:2004wa,Abe:2009dr}.
The realization of quark and lepton masses and their mixing angles was studied in Refs.~\cite{Abe:2012fj,Abe:2014vza,Fujimoto:2016zjs,Hoshiya:2021nux}.
Furthermore, their flavor structure is controlled by modular symmetry \cite{Kobayashi:2018rad,Kobayashi:2018bff,Ohki:2020bpo,Kikuchi:2020frp,Kikuchi:2021ogn,Almumin:2021fbk,Tatsuta:2021deu}.
Thus, compactification with magnetic background fluxes is 
one of practical methods to derive
realistic particle physics from string theory.

4D low energy effective theories have often been constructed
through compactification of higher-dimensional super Yang-Mills (SYM)
theory with the canonical kinetic term
\cite{Cremades:2004wa,Abe:2012ya}.
On the other hand, the Dirac-Born-Infeld (DBI) action \cite{Fradkin:1985qd,Leigh:1989jq} with the Chern-Simons (CS) terms \cite{Douglas:1995bn,Green:1996dd,Cheung:1997az,Morales:1998ux,Stefanski:1998he,Scrucca:1999uz,Scrucca:1999jq}
describes the dynamics of massless open string modes
on the D-branes.
At the lowest order of the gauge field strength $F$,
the DBI action reduces to Yang-Mills theory.
However, the DBI action can describe more stringy D-brane natures, e.g. T duality.
For non-Abelian DBI action,
higher order terms of the gauge field strength
are less-known owing to its noncommutativity
\cite{Brecher:1998su,Brecher:1998tv,Garousi:1998fg,Tseytlin:1999dj,Hashimoto:1997gm,Myers:1999ps,Johnson:2000ch}
and it is also less known
to compute explicitly 4D effective theories
via compactification on a magnetized torus.
This naturally motivates us to study
dimensional reduction of the non-Abelian DBI action 
for including higher order corrections.
Our purpose in this paper is to compute 4D $\mathcal{N}=1$
supersymmetric effective action from ten-dimensional (10D) non-Abelian DBI action compactified on the magnetized six-dimensional torus with focus on terms up to $\mathcal{O}({F}^4)$:
\begin{align}
 \mc{L}_{\text{4D}} =
 \int_{\mbb{T}^6} d^6 y \, \mc{L}_{\text{non-Abelian DBI}}
 \sim 
 \int_{\mbb{T}^6} d^6 y \, (\tr {F}^2 + \tr {F}^4)
 \quad \text{where } \hat{{F}}_{y^i y^j} \neq 0 .
\end{align}
Here $y^i~(i=1,2, \ldots , 6)$ denote the coordinates 
in the extra six dimensions and $\hat{F}$ is the background flux.
Hereafter, we drop the Neveu–Schwarz-Neveu–Schwarz (NSNS) two-form potential for simplicity throughout this paper. 
We ignore also CS terms in the D-brane action
since they mainly contribute to topological terms
and supersymmetry (SUSY) breaking terms
which vanish for supersymmetric vacua with canceled tadpoles.
We focus on bosonic part of non-Abelian DBI action in this paper, since fermions can be naturally introduced with SUSY.

In 4D action, we show the matter K\"{a}hler metric,
gauge kinetic function, and superpotential
in supergravity (SUGRA) through a systematic study of dimensional reduction.
The DBI correction of $\mathcal{O}({F}^4)$ contributes
only to the matter K\"{a}hler metric and gauge kinetic function.
It turns out that there exists a new flux contribution to 
the matter K\"{a}hler potential, 
while gauge kinetic functions and
holomorphic Yukawa couplings in the superpotential 
are consistent with previous works.
Such a new flux correction to the K\"{a}hler metric has been often neglected,
although a flux contribution to gauge coupling is
frequently discussed for the coupling unification.
We take flux corrections into account consistently in this sense and show a concrete dependence on fluxes in the K\"ahler potential of chiral matters. Also, that of open string moduli, which was discussed in Refs.~\cite{Lust:2004cx,Lust:2004fi,Font:2004cx}, is shown in Appendix~\ref{app:details}.
Such consistent treatment may become important to study swampland conjectures \cite{Vafa:2005ui}
with effective field theories (see \cite{Palti:2019pca} for a review). 
The new matter K\"{a}hler metric is independent of flavor
but depends on the fluxes, 4D dilaton, K\"{a}hler moduli and complex structure moduli,
and will be always positive definite if an induced Ramond-Ramond (RR) charge of the D-branes on which matters are living are positive.
The contribution of the matter K\"{a}hler metric to the scalar potential is shown to be consistent with the SUGRA formulation, and the superpotential is read from scalar quartic interaction.

The paper is organized as follows.
In Sec.~\ref{sec:NDBI},
we give a brief review of the non-Abelian DBI action and a magnetized torus.
In Sec.~\ref{sec:U(3)ontori},
we derive 4D supersymmetric low energy effective action from
the DBI action compactified on a magnetized torus.
The results turn out to be consistent with 4D SUGRA formulation.
Sec.~\ref{sec:summary} is devoted to the summary and discussion.
In Appendixes \ref{app:details} and \ref{app:WABC}, we give the details of the calculations.

\section{Non-Abelian DBI action on magnetized extra dimensions}
\label{sec:NDBI}

In this section, we introduce the DBI action and summarize our setup of flux compactification of the DBI action on a six-dimensional torus.

The dynamics of massless open string modes 
on the D$p$-brane is described by the DBI action with the CS terms.
The DBI action for Abelian gauge theory is expressed as
\begin{align}
 S_\DBI [ g_{MN}, \varphi, A_M] = - T_p \int d^{p+1} \xi \, e^{-\varphi} \sqrt{ -\det_{p+1}\bigl( g_{MN} 
 + 2 \pi \ap F_{MN} \bigr) },
 \label{eq:DBI}
\end{align}
where $M, N = 0,1, \ldots , p$ stand for the indices of the $(p+1)$-dimensional world volume of D$p$-brane,
and $g_{MN}$ is the pull back of the bulk metric on the D-brane.
$\ap$ denotes the Regge slope, and
$F_{MN}$ is the gauge field strength on the D$p$-brane,
$F_{MN} = \del_M A_N - \del_N A_M$.
$\varphi$ denotes the 10D dilaton field and
$T_p$ is the brane tension given by $T_p = 2\pi / \ls^{p+1} = 2 \pi / ( 2 \pi \ap^{1/2})^{p+1}$, where
$\ls = 2\pi \ap^{1/2}$ is the string length.
The superpartner fermions are dropped here for simplicity.
The DBI action \eqref{eq:DBI} is known to be robust for an Abelian gauge theory living on a single D-brane.

A non-Abelian gauge theory is realized on a stack of D-branes.
The author in Ref.~\cite{Tseytlin:1997csa} proposed the non-Abelian version of the DBI action 
with a prescription of the symmetrized trace,
while terms higher than ${\cal O}({F}^6)$ in the non-Abelian DBI (NDBI) action
are still ambiguous owing to 
its noncommutativity \cite{Brecher:1998su,Brecher:1998tv,Garousi:1998fg,Tseytlin:1999dj,Hashimoto:1997gm,Myers:1999ps,Johnson:2000ch}.
As the extension of Eq.~\eqref{eq:DBI},
NDBI action is given by \cite{Tseytlin:1997csa}
\begin{align}
 S_{\NDBI} = - T_p \int d^{p+1} \xi\,
 e^{-\varphi} \str \sqrt{ - \det_{p+1} ( g_{MN} + 2 \pi \ap F_{MN} )}.
 \label{eq:NDBI}
\end{align}
Here $F_{MN}$ is the field strength of non-Abelian gauge field, 
$F_{MN} = \del_M A_N -\del_N A_M +i[A_M , A_N]$,
and ``$\str$'' denotes the symmetrized trace,
\begin{align}
 \str ( T_1 \cdots T_n ) = \frac{1}{n!} \tr \bigl[
 T_1 \cdots T_n + (\text{permutations}) \bigr].
\end{align}

Hereafter, we consider space-filling D9-branes ($p=9$) for concreteness because the Lagrangian in the bosonic part consists only of the gauge field.
We focus on terms up to ${\cal O}({F}^4)$.

\subsection{Magnetized D9-branes on the six-dimensional torus}

We introduce background fluxes on a stack of D9-branes compactified on a six-dimensional torus.
Let us consider a six-dimensional torus consisting of
three two-dimensional tori as the extra dimension $\mbb{T}^6 = \prod_{i=1}^3 \mbb{T}^2_i$.
The 10D metric of $M_4 \times \prod_{i=1}^3 \mbb{T}^2_i$ is given by
\begin{align}
 ds_{10}^2 = e^{2\Phi} \eta_{\mu \nu} dx^\mu dx^\nu + \ls^2 
 \sum_{i=1}^3 g_{mn}^{(i)} dy_i^{m} dy_i^{n},
 \quad
 g^{(i)}_{mn} = e^{2\sigma_i}
 \begin{pmatrix}
 1 & \tau^{(i)}_R \\
 \tau^{(i)}_R & |\tau^{(i)}|^2
 \end{pmatrix},
 \label{metric}
\end{align}
where $\mu , \nu = 0,1,2,3$, $\eta_{\mu \nu} = \diag(-1,1,1,1)$ is the Minkowski metric
and $\tau^{(i)} = \tau^{(i)}_R + i\, \tau^{(i)}_I~(i=1,2,3)$ 
is the complex structure modulus on the $i$th torus $\mathbb{T}^2_i$. 
$y^m_i~(m=1,2)$ denotes the coordinate 
on $\mathbb{T}^2_i$ and $0 \leq y_i^m \leq 1$, where $y$'s are normalized by
the string length.
The volume of the $i$th torus in the string length unit 
reads
\begin{align}
 \Vol(\mbb{T}^2_i)= \sqrt{g^{(i)}} = \mc{A}^{(i)} = e^{2 \sigma_i} \tau^{(i)}_I.
\end{align}
Hence, $e^{2 \sigma_i}$ is regarded as a volume modulus of $\mathbb{T}^2_i$.
For the 4D Einstein frame, we have introduced the 4D dilaton $\Phi$,
\begin{align}
 \Phi = \varphi - \frac{1}{2} \log \prod_i \mc{A}^{(i)}
 = \varphi - \frac{1}{2} \log \Vol(\mathbb{T}^6),
\end{align}
where $ \Vol(\mathbb{T}^6) = \mc{A}^{(1)}\mc{A}^{(2)}\mc{A}^{(3)}$ is the volume of
$\mathbb{T}^6$.
With the complex coordinate on the $i$th torus
\begin{align}
 dz_i = dy_{i}^1 + \tau^{(i)} dy_{i}^2,
 \quad
 i = 1 , 2, 3,
\end{align}
the 10D metric is rewritten as
\begin{align}
 ds_{10}^2 = e^{2\Phi} \eta_{\mu \nu} dx^\mu dx^\nu 
 + \ls^2 \sum_{i=1}^3 e^{2 \sigma_i} dz_i d\overline{z_i} . \label{eq:metric}
\end{align}
Thus, the metric on the $\mbb{T}^2_i$ in the the complex basis is given by
\begin{align}
 g_{i\bar{j}} = \ls^2 \frac{e^{2 \sigma_i}}{2} \delta_{i\bar{j}}.
 \label{eq:metricoftorus}
\end{align}

\medskip

We shall focus on a stack of the space-filling D9-branes on
the factorized torus $\prod_{i=1}^3 \mbb{T}^2_i$
with nontrivial background fluxes
on the D-branes.
The NDBI action \eqref{eq:NDBI} expanded up to $\mathcal{O}({F}^4)$ is given by 
\cite{Tseytlin:1997csa}
\begin{align}
 S_\NDBI \approx &~ - T_9 \int d^{10} X 
 \sqrt{-\det  g_{MN}}
 \, e^{-\varphi}
 \frac{(2\pi \ap)^2}{4} \tr \biggl[ {F}_{MN} F_{MN}
 - \frac{(2\pi\ap)^2}{3}\biggl( {F}_{KL} {F}_{LM} {F}_{NK} {F}_{MN} 
 \nn \\
 &+ \frac{1}{2} {F}_{KL} {F}_{LM} {F}_{MN} {F}_{NK} -\frac{1}{4} {F}_{KL} {F}_{KL} {F}_{MN} {F}_{MN}
 - \frac{1}{8} {F}_{KL} {F}_{MN} {F}_{KL} {F}_{MN}\biggr)
 + \mc{O}({F}^6)
 \biggr]
 ,
  \label{eq:NDBI0}
\end{align}
where the metric is omitted in contracting indices of the gauge field strength, e.g.
${F}_{MN} {F}_{MN} \coloneqq g^{MK} g^{NL} {F}_{MN} {F}_{KL}$.
$X$ denotes the bulk coordinate in 10D.
The normalization of gauge group generator is
assumed to be given by $\tr (T^a T^b) =\delta^{ab}$. 
The quadratic term $\tr{F}_{MN}^2$
can reduce to the well-known Yang-Mills action with the canonical kinetic term.

With respect to the background fluxes on the D9-branes,
it is assumed that only the fluxes on the extra six dimension have nonzero values,
\begin{align}
 {F}_{MN} \ni 
 \hat{F}_{y^m_i y^n_j},
 \quad \text{where}~
 \hat{F}_{y^1_i y^2_i} \neq 0
 .
\end{align}
Here,
the background flux $\hat{F}$ is taken to be
diagonal with respect to the torus index, i.e.,
$\hat{F}_{y^1_i y^2_i} \neq 0$ for $i=1,2,3$
and $\hat{F}_{y^m_i y^n_j} = 0$ for $i\neq j$.
In the complex basis,
nonvanishing components of the fluxes are given by
\begin{align}
 \hat{F}_{z_i \bar{z}_i}
 = \frac{\del y_i^m}{\partial z_i} 
 \frac{\partial y_i^n}{\partial \bar{z}_i} 
 \hat{F}_{y^m_i y^n_i},~~~
 i=1,2,3.
\end{align}
See Appendix~\ref{app:details} for details.
This is consistent with the SUSY condition
as discussed later.

\subsection{Flux and matter zero modes}
\label{sec:zero-mode}

Although fermions are neglected so far,
they exist in the presence of the SUSY. 
We briefly review a zero (massless) mode solution 
of the Dirac equation on the 
$\mathbb{T}^2$ with $U(1)$ magnetic flux \cite{Cremades:2004wa}.
A generalization of the solution to the $\mathbb{T}^6$ case is discussed later.

The background magnetic flux on $\mathbb{T}^2$ 
in the string length unit
is given by
\begin{align}
 \int_{\mbb{T}^2} \frac{\hat{F}}{2\pi} = M ~ \to~
 \hat{F} = \frac{\pi i M}{\tau_I  } dz \wedge d\bar{z}
 ,\quad M \in {\mathbb Z}.
\end{align}
Then, the gauge potential can be written as
\begin{align}
 \hat{A}(z) = \frac{\pi M}{\tau_I } \Im (\bar{z} \, dz).
\end{align}
A large gauge transformation associated with translations on the
torus is given by
\begin{align}
 \hat{A}(z+1) = 
 \hat{A}(z) + 
 d \bigg( \frac{\pi M}{\tau_I } {\rm Im}(z) \bigg) , \quad
 \hat{A}(z+\tau) = 
 \hat{A}(z) + 
 d \bigg( \frac{\pi M}{\tau_I } {\rm Im}(\bar{\tau}z) \bigg).
\end{align}
Let us consider the spinor $\psi$ on the $\mathbb{T}^2$ with unit charge $q=1$,
where
\begin{align}
 \psi = 
 \begin{pmatrix}
 \psi_+ (z) \\
 \psi_- (z)
 \end{pmatrix}
 .
\end{align}
Here, $\pm$ denotes the eigenvalue of $SO(2)$ 
spinor algebra associated with the torus (chirality).
The gauge transformation acts on the spinor as
\begin{align}
 \psi(z+1) = 
 \exp\bigg[i \frac{\pi M}{\tau_I } {\rm Im}(z)\bigg]\psi(z) ,
 \quad
 \psi(z+\tau) = 
 \exp\bigg[i  \frac{\pi M}{\tau_I } {\rm Im}(\bar{\tau}z)\bigg]\psi(z) .
\end{align}
With these two boundary conditions, we solve
the Dirac equation 
$i \slashed{D}\psi = 0$ on
the $\mathbb{T}^2$. It is noted that
the spinor becomes a single-valued function up to the gauge transformation
when $M \in {\mathbb Z}$.
For $M>0$,
only $\psi_+$ is a normalizable zero modes, which is $|M|$-fold degenerate;
similarly, only $\psi_-$ is a normalizable $|M|$-fold degenerate zero mode for $M< 0$.
Hence, the effective theory becomes chiral in
the low energy limit.
Explicitly, for $M>0$ the $\psi_+$ is written as   
\begin{align}
 \psi_{{\mathbb T}^2}^{A,M} =
 \Theta^{A,M}(z) \coloneqq
 \mc{N}_M
 \exp\bigg[\pi i M z \frac{{\rm Im}(z)}{\tau_I}\bigg]
 \vartheta
 \begin{bmatrix}
 \frac{A}{M} \\
 0
 \end{bmatrix}
 (Mz, M\tau),
 \quad
 A = 0,1, \cdots, M-1 .
 \label{eq:zero-f1}
\end{align}
Here, $\mc{N}_M$ is the normalization constant,
$A$ labels the number of degeneracy, i.e. flavor,
and
$\vartheta$ is the Jacobi theta function
\begin{align}
 \vartheta
 \begin{bmatrix}
 a \\
 b
 \end{bmatrix}
 (\nu, \tau) 
 \coloneqq \sum_{l \in {\mathbb Z}} 
 e^{\pi i (a+l)^2\tau} e^{2\pi i (a+l)(\nu +b)}.
\end{align}
The normalization of $\psi_+$ reads
\begin{align}
 \int_{\mbb{T}^2} d^2y \sqrt{g_2}\,
 \ol{\Theta^{A,M}(z) } 
 \Theta^{B,M}(z)
 = \delta^{AB}
 \frac{(\mc{N}_M)^2 \mc{A}}{\sqrt{2\tau_I |M|}},
\end{align}
and we choose the following condition\footnote{
The normalization factor for $M=0$ is $\mc{N} = 1 / \sqrt{ \mc{A}}$.
}
\begin{align}
 (\mc{N}_M)^2 = \frac{\sqrt{2 \tau_I |M|}}{\mc{A}}
 \label{eq:normalization-1}
\end{align}
such that $\int_{\mbb{T}^2} d^2y \sqrt{g_2}\,
 \ol{\Theta^{A,M}(z) } 
 \Theta^{B,M}(z) = \delta^{AB}$.
Here, we used dimensionless coordinate $z$.
For $M<0$, the normalizable solution of $\psi_-$ is written as
\begin{align}
\psi_{{\mathbb T}^2}^{A,M}
= \overline{\Theta^{A,|M|}(z) } , \quad
A= 0, 1, \ldots , |M|-1,
\end{align}
where the normalization constant is the same as that for $M>0$.
Thus, a signature of $M$ is associated with 
the chirality of fermion.

\subsubsection{Symmetry breaking of $U(3) \to U(1)_a \times U(1)_b \times U(1)_c$ and degeneracy}

It is easy to extend the above solution to a 10D theory 
compactified on
${\mathbb T}^6=\prod_{i=1}^3 \mbb{T}^2_i$
with non-Abelian gauge symmetries of our interest.
In the 10D SYM theory, there exist gauge fields $A_M$ and 
their superpartner gluinos $\lambda^{(10)}$.

It is necessary to take into account of background fluxes
to identify which zero modes survive in 4D theory.
We give the following background fluxes in a non-Abelian gauge theory:
\begin{align}
 \hat{F}_{z_i \bar{z}_i}
 \eqqcolon
  \hat{F}_{i \bar{i}}
= \frac{i \pi M^{(i)}}{\tau_I^{(i)} },\quad i=1,2,3.
\end{align}
Here, $M^{(i)}$ is a matrix-valued constant,
and gives the gauge symmetry which can survive in the 4D theory through $[M^{(i)}, A_\mu]=0$.
Otherwise, gauge fields become massive.
For simplicity, we hereafter 
focus on the case in which 
the $U(3)$ gauge group in 10D is broken to 
$U(1)_a \times U(1)_b \times U(1)_c$ in 4D by
the diagonal background fluxes,
\begin{align}
  \frac{1}{2\pi} \int_{\mbb{T}^2_i} 
  dz_i \wedge d\bar{z}_i \, \hat{F}_{i \bar{i}}=
  M^{(i)} = \begin{pmatrix}
  \fM{i}{a} & & \\
  & \fM{i}{b} & \\
  & & \fM{i}{c}
  \end{pmatrix},
  \quad
  \fM{i}{a,b,c} \in \mbb{Z},
  \label{eq:backgroundflux}
\end{align}
where the fluxes are similarly quantized
for a charged zero mode 
to have a single-valued function on the each $\mathbb{T}^2$ 
up to gauge transformation. 
Replacing unity with identity matrix
in Eq.~\eqref{eq:backgroundflux} can realize
4D non-Abelian gauge symmetries. 
It is noted that gauge fields and gluinos in 10D are both adjoint representations, in which they are coupled to the fluxes
with a commutator through their covariant derivatives.
Hence, the degeneracy of fermion zero modes $ I_{\alpha \beta}$ depends on 
the difference of fluxes between two gauge groups
on each torus \cite{Cremades:2004wa},
\begin{align}
 I_{\alpha \beta} \coloneqq \prod_{i=1}^3 \fI{i}{\alpha \beta}, 
 \qquad
 \fI{i}{\alpha \beta} \coloneqq M_{\alpha}^{(i)} - M_{\beta}^{(i)}
 \quad
 (\alpha,~\beta = a,~b,~c),
\end{align}
for matter with a charge of $(1,-1)$ 
against a $U(1)_{\alpha} \times U(1)_{\beta}$ gauge group.
It is noted that the definition of $I_{\alpha \beta}$ gives
\begin{align}
 I_{ab}^{(i)} + I_{bc}^{(i)} + I_{ca}^{(i)} =0.
 \label{eq:gauge-singlet}
\end{align}
This equation can determine a relative signature 
among $I_{\alpha \beta}$'s.
Next, we show the SUSY condition for avoiding tachyons,
and visit concrete zero mode functions.

\subsubsection{Supersymmetry conditions on the background fluxes}

We consider the condition for background fluxes 
to preserve 4D $\mc{N}=1$ SUSY
for realizing chiral theories.
The SUSY transformation of 10D fermions
should vanish to preserve the 4D SUSY.
Then, the condition of the background fluxes 
at $\mc{O}(\hat{F}^2)$
on the complex manifold reads 
\cite{Becker:1995kb,Marino:1999af,Lust:2004cx,Cremades:2004wa,Haack:2006cy,Blumenhagen:2006ci}:
\begin{align}
 &g^{i\bar{j}} \hat{F}_{i \bar{j}} = 0,
  \label{eq:SUSY3-1}
 \\
 & \hat{F}_{i j} = \hat{F}_{\bar{i} \bar{j}}=0.
 \label{eq:SUSY3-2}
\end{align}
It is noted that an additional
term of
$\hat{F}_{z_1 \bar{z}_1} \hat{F}_{z_2 \bar{z}_2} \hat{F}_{z_3 \bar{z}_3}$
to the rhs of Eq.~\eqref{eq:SUSY3-1}
is required for the calibration condition of magnetized D-branes with DBI action. 
However, the above condition is sufficient to us
since we focus on the terms
of $\mathcal{O}(F^4) \ni \hat{F}^2 \times (\mr{fluctuations})$ 
in the Lagrangian.
Higher order corrections in 
$\mc{O}(F^6)  \ni \hat{F}^4 \times (\mr{fluctuations})$ neglected 
in this paper can modify the condition to
the terms involved in $F^4$.
In our case, the former condition \eqref{eq:SUSY3-1}
is satisfied when
\begin{align}
\sum_{i=1}^3 \frac{\fM{i}{\alpha}}{\mc{A}^{(i)}}
 =0.
 \qquad (\alpha = a, b, c).
 \label{eq:SUSY2}
\end{align}
The latter condition on the vanishing holomorphic flux condition \eqref{eq:SUSY3-2}
is satisfied when we consider 
the diagonal fluxes in the torus index.
Then tachyons are absent in the effective theories since
their mass squared is proportional to
\cite{Troost:1999xn,Cremades:2004wa,Abe:2008fi}
\begin{align}
 \sum_{i=1}^3 \frac{\fM{i}{\alpha} - \fM{i}{\beta}}{\mc{A}^{(i)}} 
 = \sum_{i=1}^3 \frac{\fI{i}{\alpha \beta}}{\mc{A}^{(i)}} =0 .
 \label{eq:SUSY}
\end{align}
This equation can also determine a relative signature 
among $I_{\alpha \beta}^{(i)}$'s with fixed $\alpha$
and $\beta$ on top of Eq.~\eqref{eq:gauge-singlet}.
For later convenience,
we introduce the notation of the flux divided by the torus area as
\begin{align}
 \fm{i}{\alpha} \coloneqq \frac{\fM{i}{\alpha}}{\mc{A}^{(i)}}
 \qquad (\alpha = a, b, c).
 \label{eq:flux-mod}
\end{align}

\subsubsection{Matter zero modes in SUSY theories}

We shall consider zero mode functions on ${\mathbb T}^6$ in the presence of 4D SUSY.
Let us take 10D chirality of the gluino $\lambda^{(10)}$ as \cite{Abe:2012ya}
\begin{align}
 \Gamma \lambda^{(10)} = + \lambda^{(10)}.
\end{align}
Then, the gluino is decomposed into the irreducible spinor representation with $SO(2)^3$ that is the Cartan subalgebra of $SO(6)$,
\begin{align}
 \lambda_0 \coloneqq \lambda_{+++},~~~
 \lambda_1 \coloneqq \lambda_{+--},~~~
 \lambda_2 \coloneqq \lambda_{-+-},~~~
 \lambda_3 \coloneqq \lambda_{--+},
 \label{eq:gaugino-dec}
\end{align}
where $\pm$ denotes the eigenvalues of $SO(2)^3$ 
spinor algebra (chiralities).
10D gauge fields $A_M$ can be decomposed similarly into
\begin{align}
 A_\mu ,~~~
 A_{z_1}  ,~~~
 A_{z_2},~~~
 A_{z_3},
\end{align}
where $A_{z_i} = \frac{i}{2\tau_I^{(i)}}(\bar{\tau^{(i)}} A_{y_i^1} - A_{y_i^2})$.
In 4D ${\cal N}=1$ SUSY theories,
a vector multiplet $V$ consists of
$A_\mu$ and $\lambda_0$, whereas 
chiral multiplets $\Phi_i$ 
can consist of fluctuations of $A_{z_i}$ 
and $\lambda_i~(i=1,2,3)$.
When the background fluxes preserve the 4D SUSY in flat spacetime,
bosonic partners have the same zero mode function as fermions' \cite{Cremades:2004wa,Abe:2012ya}.
Then, the zero mode function of the massless gauge multiplet $V$ is independent of coordinates $y$ since there exists no coupling to the fluxes
in the zero mode equation, i.e., $[M^{(i)}, A_\mu]=0$.
For the chiral multiplets $\Phi_i(x)$, the zero mode functions $\phi_i(y)$
are given by
products of those on each torus:
\begin{align}
 \Phi_i^{\text{10D}}
 (x,y) &= \sum_{{\mathbb A}} 
 \Phi_i^{{\mathbb A}, I_{\alpha \beta}} (x) \otimes 
 \phi_{i}^{{\mathbb A}, I_{\alpha \beta}}(y)
 ~+~({\rm massive~modes}), \\
 \phi_{i}^{{\mathbb A}, I_{\alpha \beta}}(y)
 &= \bigg(\prod_{r=1}^3 
 \phi_{i, {\mathbb T}_r^2}^{A^{(r)}, I_{\alpha \beta}^{(r)}}(y_r)
 \bigg).
 \label{eq:zero-f2}
\end{align}
Here, for $I_{\alpha \beta} \neq 0$, 
\begin{align}
 \phi_{i, {\mathbb T}_r^2}^{A^{(r)}, I_{\alpha \beta}^{(r)}}
 =
 \begin{cases}
 \Theta^{A^{(r)}, I_{\alpha \beta}^{(r)}}(z_r) 
 ~{\rm with}~ \tau^{(r)} & 
 (r=i ~\&~ I_{\alpha \beta}^{(r)} >0 ), \\
 \overline{\Theta^{A^{(r)}, |I_{\alpha \beta}^{(r)}|}(z_r) } ~{\rm with}~ \ol{\tau^{(r)}} & 
 (r \neq i~\& ~ I_{\alpha \beta}^{(r)} <0), \\
 0 & ({\rm other ~cases}).
 \end{cases}
 \label{eq:zero-f3}
\end{align}
This is consistent with chiralities in Eq.~\eqref{eq:gaugino-dec}.
$A^{(r)}$ is the index of flavor on each torus:
$A^{(r)} = 0,1,\ldots |I_{\alpha \beta}^{(r)}|-1$
,and hence, the total flavor index is ${\mathbb A}= 0 , 1, \ldots , |I_{\alpha \beta}| -1$. 
It is noted that matter $ \Phi_i^{{\mathbb A}, I_{\alpha \beta}}$
has a charge of $(1,-1)$ 
against the $U(1)_{\alpha} \times U(1)_{\beta}$ gauge group.

Without loss of generality, we assume that
\begin{align}
 & I_{ab}^{(1)} >0, \qquad I_{ab}^{(2),(3)} <0, \nn \\
 & I_{bc}^{(2)} >0, \qquad I_{bc}^{(1),(3)} <0, 
 \label{eq:signature}
 \\
 & I_{ca}^{(3)} >0, \qquad I_{ca}^{(1),(2)} <0, \nn
\end{align}
to satisfy Eqs.~\eqref{eq:gauge-singlet} and
\eqref{eq:SUSY}.
This is also consistent with decomposition of Eq.~\eqref{eq:gaugino-dec} as below.
As noted, we have the gauge symmetry breaking of
$U(3) \to U(1)_a \times U(1)_b \times U(1)_c$.
Then, the fluctuations of 10D gauge fields are decomposed
into 4D zero (massless) modes,
which are, namely, gauge fields 
and complex scalars charged under the 4D gauge symmetries:
\begin{align}
 & a_\mu = \begin{pmatrix}
 a^a_\mu & &  \\
 & a^b_\mu &  \\
 & & a^c_\mu 
 \end{pmatrix},
 \label{eq:gaugefields}
 \\
 &
a_{z_i}
= \begin{pmatrix}
 & a^{ab}_i & \\
 & & a^{bc}_i \\
 a^{ca}_i & &
 \end{pmatrix}
 \eqqcolon \begin{pmatrix}
 & A_i \phi_i^{ab} \delta_{i1} & \\
 & & B_i \phi_i^{bc} \delta_{i2}  \\
 C_i \phi_i^{ca} \delta_{i3} & &
 \end{pmatrix},
 \label{eq:chargedmatters}
\end{align}
where $a_M$ denotes fluctuations of the 10D gauge fields;
$a^{a,b,c}_\mu$ are the 4D gauge fields associated with $U(1)_{a,b,c}$ symmetries, $A_i$, $B_i$ and $C_i$ 
denote 4D complex scalars.
$(\phi_i^{ab}, \phi_i^{bc}, \phi_i^{ca}): =
(\phi_i^{I_{ab}}, \phi_i^{I_{bc}}, \phi_i^{I_{ca}})$ show zero mode functions relevant to each complex scalar
and we suppressed the flavor index.
These scalars have bifundamental charges against 
$U(1)_a \times U(1)_b \times U(1)_c$ symmetries,
$Q(A_i) =  (1, -1, 0)$, $Q(B_i) = (0, 1, -1)$
and $Q(C_i) = (-1, 0, 1)$, respectively,
where $Q({\rm scalar})$ denotes the $U(1)$ charges of the scalar.
According to Eqs.~\eqref{eq:zero-f3} and \eqref{eq:signature},
the surviving zero modes in 4D are only
\begin{align}
 A_1^{\mathbb{A}} ,~B_2^{\mathbb{B}} ~\text{and}~ C_3^{\mathbb{C}}.
\end{align}
Here, $\mathbb{A},~\mathbb{B},$ and $\mathbb{C}$ are the flavor indices, and
their zero mode functions surviving in 4D 
are written as
\begin{align}
 \phi_1^{{\mathbb A}, ab} &= \Theta^{A^{(1)}, I_{ab}^{(1)}} (z_1) \otimes  \overline{\Theta^{A^{(2)}, |I_{ab}^{(2)}|} (z_2 )}
 \otimes  \overline{\Theta^{A^{(3)}, |I_{ab}^{(3)}|} 
 (z_3 )} , \nn \\
 \phi_2^{{\mathbb B}, bc} &=  
 \overline{\Theta^{B^{(1)}, |I_{bc}^{(1)}|} (z_1)} \otimes \Theta^{B^{(2)}, I_{bc}^{(2)}} (z_2 )
 \otimes  \overline{\Theta^{B^{(3)}, |I_{bc}^{(3)}|}
 (z_3 )} ,
 \label{eq:zero-function} 
 \\
 \phi_3^{{\mathbb C}, ca} &=  \overline{\Theta^{C^{(1)}, |I_{ca}^{(1)}|} (z_1)} \otimes  \overline{\Theta^{C^{(2)}, |I_{ca}^{(2)}|} (z_2 )}
 \otimes  \Theta^{C^{(3)}, I_{ca}^{(3)}} (z_3 ),
 \nn
\end{align}
where $A^{(r)} = 0 ,1, \cdots |I_{ab}^{(r)}|-1$,
$B^{(r)} = 0 ,1, \cdots , |I_{bc}^{(r)}|-1$, and
$C^{(r)} = 0 ,1, \cdots , |I_{ca}^{(r)}|-1$~($r=1,2,3$):
${\mathbb A} = 0 ,1, \cdots |I_{ab}|-1$,
${\mathbb B} = 0 ,1, \cdots , |I_{bc}|-1$, and
${\mathbb C} = 0 ,1, \cdots , |I_{ca}|-1$.
The normalization factor of $\Theta^{A^{(1)}, \fI{1}{ab}}(z_1)$ is denoted as
$\mc{N}^{1}_{\fI{1}{ab}}$ for instance.
From Eq.~\eqref{eq:normalization-1},
these zero mode functions are normalized as
\begin{align}
 \int_{\mbb{T}^6} d^6y \sqrt{g_6} \, \phi^{\mbb{A},ab}_1 \ol{\phi^{\mbb{A}',ab}_1} = \delta_{\mbb{A}, \mbb{A}'}.
 \label{eq:normalization-wavefunction}
\end{align}

$\phi$'s are zero mode solutions for 10D SYM 
with the canonical kinetic term.
In the case with the NDBI action, 
there are corrections of fluxes to this zero mode solution.
Since the flux is constant to the coordinates of a six-dimensional torus, the corrections are expected to change
the normalization of the matter K\"{a}hler metric. 
In this paper, for simplicity, we neglect higher order 
interactions with derivatives in 4D theories such as 
$|A|^2|\partial A|^2$ or $|\partial A|^4$, 
where $A$ is a 4D complex scalar in a chiral matter multiplet.

\section{SUSY effective action of
$U(1)_a \times U(1)_b \times U(1)_c$ theory}
\label{sec:U(3)ontori}

In this section,
we will exhibit 4D SUSY effective action
derived from the 10D NDBI action, focusing on the bosonic sector.
As noted already, we assume to start with 10D $U(3)$ gauge symmetry which is 
broken to $U(1)_a \times U(1)_b \times U(1)_c$
by the background flux of Eq.~\eqref{eq:backgroundflux}.

We can read the 4D gauge couplings, K\"{a}hler metrics of the chiral matters and scalar quartic couplings,
after substituting the fields of Eqs.~\eqref{eq:gaugefields}, \eqref{eq:chargedmatters} and the metric \eqref{eq:metric}
into the NDBI action \eqref{eq:NDBI0}.
For later convenience, we define closed string moduli
\cite{Cremades:2004wa}:
\begin{align}
 s 
 & \coloneqq e^{-\varphi} {\cal A}^{(1)} {\cal A}^{(2)} {\cal A}^{(3)} 
 = e^{-\varphi}{\rm Vol}(\mathbb{T}^6),
 \\
 t_i 
 & \coloneqq e^{-\varphi} {\cal A}^{(i)} = e^{-\varphi}{\rm Vol}(\mathbb{T}_i^2) ,
 \\
 U_i  
 & \coloneqq i \overline{\tau^{(i)}},\quad
 u_i \coloneqq \Re (U_i) = \tau^{(i)}_I ,
\end{align}
where $s$ is the 4D dilaton, and $t_i$ are the K\"{a}hler moduli.
$U_i$ stand for the complex structure moduli of ${\mathbb T}^2_i$ in the SUGRA basis.
In combination with axions descended from RR tensors,
the above moduli constitute the complexified dilaton $S$ and the K\"{a}hler moduli $T_i$.
The K\"{a}hler potential of these closed string moduli $K^{(0)}$ is given by
\begin{align}
 & K^{(0)} = -\log(S+\ol{S}) - \sum_{i=1}^3 \log(T_i+\ol{T_i})
 - \sum_{i=1}^3 \log(U_i+\ol{U_i}).
 \label{K-moduli}
\end{align}
4D effective action of chiral matters is written with these closed string moduli as seen below.
See Appendix \ref{app:details} for details of the computation.

\subsection{Gauge couplings}

The gauge couplings of $U(1)_a \times U(1)_b \times U(1)_c$ 
are read from the coefficient of the gauge kinetic term.
The canonical kinetic term 
$\mc{L}_{\text{4D}} \ni \int d^6 y \sqrt{g_6}\, e^{-\varphi} \tr (f_{\mu \nu})^2$ gives
the leading contribution without fluxes, whereas the flux-corrected 
contributions come from\footnote{
A contribution of 
$\int d^6 y \sqrt{g_6} \,e^{-\varphi}{\rm tr}
[\hat{F}_{i\bar{j}}f_{\mu \nu} 
{\hat{F}}^{\bar{j}i} f^{\mu \nu}]$
is included because 
$[\hat{F}_{i\bar{j}},f_{\mu \nu} ] =0.$
}
$\mc{L}_{\text{4D}} \ni \int d^6 y \sqrt{g_6} \,e^{-\varphi}{\rm tr}[
({\hat{F}}_{i\bar{j}}{\hat{F}}^{\bar{j}i})(f_{\mu \nu})^2]$.
Here, $f_{\mu \nu}$ is the fluctuation of the 10D gauge field strength of the $U(3)$ gauge symmetry with the 4D subscripts.
The former kinetic term
depends on $e^{-\varphi}{\rm Vol}(\mathbb{T}^6) =s$ and
the latter includes $s \times m^2$, where $m$ is 
moduli-dependent flux defined in Eq.~\eqref{eq:flux-mod}.
Thus, we find
\begin{align}
 S_\NDBI \ni - \frac{1}{2\pi} \int d^4x \sqrt{-g_4}\, \frac{1}{4g_a^2} (f^{a}_{\mu\nu})^2,
\end{align}
where $f_{\mu \nu}^a = \del_\mu a_\nu^a - \del_\nu a_\mu^a$ 
is the field strength for the $U(1)_a$, and
the gauge coupling for the $U(1)_a$ group is 
\begin{align}
 \frac{1}{g_a^2} =&~ s \biggl[ 1 + \frac{1}{2} \sum_{i=1}^3 (m^{(i)}_a)^2 \biggr]
 \\
 =&~ s -t_1 \fM{2}{a} \fM{3}{a} - t_2 \fM{1}{a} \fM{3}{a} - t_3 \fM{1}{a} \fM{2}{a}.
\end{align}
In the second line, the SUSY condition \eqref{eq:SUSY2} is used.
The results for $U(1)_b$ and $U(1)_c$ symmetries are similar to that of the $U(1)_a$. 
This is a well-known result of the D-brane models \cite{Lust:2004cx,Lust:2004fi,Font:2004cx,Blumenhagen:2006ci}
and is regarded as the real part of a corresponding holomorphic gauge coupling $f_a$,
\begin{align}
 \Re(f_a) &= \frac{1}{g_a^2}, \\
 f_a &= S -T_1 \fM{2}{a} \fM{3}{a} - T_2 \fM{1}{a} \fM{3}{a} - T_3 \fM{1}{a} \fM{2}{a} . 
 \label{eq:hol-gauge}
\end{align}
The expansion in fluxes is valid when $s > t_i |M^{(j)}_aM^{(k)}_a|~(i \neq j \neq k \neq i)$.
Then a gauge coupling will become weak for large vacuum expectation values of moduli.
It is noted that terms dependent on $T_i$ can be positive contributions to the gauge coupling
when an induced D5-brane charge $-M^{(j)}_a M^{(k)}_a$, which is carried by a magnetized D9-brane, is positive.\footnote{
The induced charge and its contribution to a holomorphic gauge coupling are seen from CS term on a D9-brane, 
$\int_{D9} ( C_6 + \frac{1}{2} C_2 \wedge f \wedge f) \wedge \hat{F} \wedge \hat{F} $, where $C_2$ and $C_6$ are RR two-from and six-form potential.
}

\subsection{K\"{a}hler metric of chiral matters}

The coefficient of a scalar kinetic term
gives the K\"{a}hler metric for chiral matter
in SUSY theories. 
The kinetic terms with the leading contribution without 
fluxes are read from
$\mc{L}_{\text{4D}} \ni
\int d^6y\,  \sqrt{g_6} e^{2\Phi -\varphi}
\tr(f_{\mu i}f^{\mu i})$,
whereas the next leading contributions with fluxes
are roughly given by a combination of
$\mc{L}_{\text{4D}} \ni
\int d^6y\sqrt{g_6} \, e^{2\Phi -\varphi}
\tr (\hat{F}_{j \bar{k}}\hat{F}^{\bar{k}j } 
f_{\mu i}f^{\mu i} +
\hat{F}_{j \bar{k}} f_{\mu i}\hat{F}^{\bar{k}j}
f^{\mu i} )$ and similar terms. Here, 
$f_{\mu i}\coloneqq f_{\mu z^i}$ 
is the fluctuation of 10D field strength
and includes the 4D kinetic term of 
a scalar fluctuation, e.g. $\del_\mu A^{\mbb{A}}_i$,
where $A_i$ is given in Eq.~\eqref{eq:chargedmatters}
with the intersection number \eqref{eq:signature}.\footnote{
We have generalized $A_1$ to $A_i$ with any $i$. 
}
A factor $e^{2\Phi}$ originates from 
the 4D Einstein frame metric
$g_{\mu \nu} = e^{-2\Phi}\tilde{g}_{\mu \nu}$ in the kinetic term
$\sqrt{-\tilde{g}_4}\times 
\tilde{g}^{\mu \nu} g^{i\bar{i}}
f_{\mu i}f_{\nu \bar{i}}$, where
$\tilde{g}_{\mu \nu}$ is the Jordan frame metric,  
$ds_{10}^2 \ni  e^{2\Phi} g_{\mu \nu} dx^\mu dx^\nu 
\eqqcolon  \tilde{g}_{\mu \nu} dx^\mu dx^\nu $ as in Eq.~\eqref{metric}.
For instance, we roughly estimate
\begin{align}
 \int d^6y\,  \sqrt{g_6} e^{2\Phi -\varphi}
 \tr(f_{\mu i}f^{\mu i})
 &\sim
 e^{2\Phi-\varphi} g^{i\bar{i}}|\del_\mu A^{\mbb{A}}_i|^2 
 \int d^6 y \sqrt{g_6} \, |\phi_i^{\mbb{A}, ab}|^2 \nn \\
 &\sim 
 \frac{2u_i}{t_i \Vol(\mbb{T}^6)} 
 |\del_\mu A^{\mbb{A}}_i|^2,
\end{align}
for terms without fluxes and
\begin{align}
 \int d^6y\,  \sqrt{g_6} e^{2\Phi -\varphi}{\rm tr}
 (\hat{F}_{j \bar{k}}\hat{F}^{\bar{k}j } 
 f_{\mu i}f^{\mu i}) 
 &\sim
 e^{2\Phi-\varphi}
 \hat{F}_{j \bar{k}}\hat{F}^{\bar{k}j}
 g^{i\bar{i}} |\del_\mu A^{\mbb{A}}_i|^2 
 \int d^6 y \sqrt{g_6}\, 
 |\phi_i^{\mbb{A}, ab}|^2 \nn \\
 &\sim 
 m^2 \times \frac{2u_i}{t_i \Vol(\mbb{T}^6)} |\del_\mu A^{\mbb{A}}_i|^2,
\end{align}
for the flux-corrected terms
with the moduli-dependent fluxes $m$ in Eq.~\eqref{eq:flux-mod}.
Here, $\phi_i^{\mbb{A},ab}$ is the zero mode function
for $A_i$ in the magnetized extra dimension,
and we used $g^{i\bar{i}} = 2e^{-\varphi}\frac{u_i}{t_i}$ and the normalization of $\phi_i^{\mbb{A},ab}$ in Eq.~\eqref{eq:normalization-wavefunction}.
In addition,
let us rescale the matter field as $A_i \to \alpha_{ab}^{(i)} A_i$
so that matter superpotential becomes a holomorphic function of the moduli and the matter K\"{a}hler metric results in a real function of the moduli \cite{Abe:2012ya}, where
\begin{align}
 \alpha^{(i)}_{\alpha \beta} = \frac{1}{\sqrt{2^2 u_i}}
 \frac{\sqrt{ \Vol(\mbb{T}^6)}}{(2^3 u_1 u_2 u_3)^{1/4}}
 \biggl( \frac{|\fI{i}{\alpha \beta}|}{\prod_{r \neq i } |\fI{r}{\alpha\beta}|} \biggr)^{1/4}, ~~~
 \alpha, \beta = a,b,c,
\end{align}
for $I_{ab} I_{bc} I_{ca} \neq 0$.
Then, the metric for $A_i$, $\mc{Z}^{i}_{ab}$, 
is obtained as
\begin{align}
  S_\NDBI \ni - \frac{1}{2\pi}\int d^4x \sqrt{-g_4}\, \mc{Z}^{i}_{ab} |D_\mu A^{\mbb{A}}_i|^2 ,
\end{align}
where $D_\mu A^{\mbb{A}}_i = (\del_\mu + i a_\mu^a -i a_\mu^b)A^{\mbb{A}}_i$, and
\begin{align}
 \mc{Z}^{i}_{ab}
 =& Z^{i}_{ab} \times \biggl[
 1 - \frac{1}{6} \bigl(
 2 \fm{j}{a} \fm{k}{a} + 2 \fm{j}{b} \fm{k}{b} + \fm{j}{a} \fm{k}{b} + \fm{j}{b} \fm{k}{a}
 \bigr) \biggr]
 \quad ( i \neq j \neq k \neq i)
 \nn \\
 =& Z^{i}_{ab} \times \biggl[
 1 - \frac{t_i}{6 s} \bigl(
 2 \fM{j}{a} \fM{k}{a} + 2 \fM{j}{b} \fM{k}{b} + \fM{j}{a} \fM{k}{b} + \fM{j}{b} \fM{k}{a}
 \bigr)
 \biggr]
 \quad ( i \neq j \neq k \neq i),
 \label{eq:KahlermetciesZ}
 \\
 Z^{i}_{ab} \coloneqq & \frac{2 u_i}{t_i \Vol(\mbb{T}^6)} \bigl( \alpha^{(i)}_{ab}\bigr)^2 
 =\frac{1}{2t_i}\biggl( \prod_{k=1}^3 \frac{1}{\sqrt{2 u_k}} \biggr) 
 \sqrt{\frac{|\fI{(i)}{ab}|}{\prod_{j\neq i}|\fI{(j)}{ab}|}}.
 \label{eq:Z}
\end{align}
Here, we used the SUSY condition in the computation.
It is noted that $Z^{(i)}_{ab}$ is the metric obtained 10D SYM with the canonical kinetic term 
on the magnetized extra dimension \cite{Cremades:2004wa,Abe:2012ya}
and that the above $\mc{Z}^{(i)}_{ab}$ in Eq.~\eqref{eq:KahlermetciesZ}
is symmetric under exchange of $a$ and $b$
and independent of labels of flavor.
This is also rewritten with complexified moduli and intersection numbers as
\begin{align}
 \mc{Z}^{i}_{ab} 
 =&
 Z^{i}_{ab} \times \biggl[ 1 + \frac{ (T_i + \ol{T_i}) }{6 ( S + \ol{S} )}
 \bigl(
 \fI{j}{ab} \fI{k}{ab} - 3 \fM{j}{a} \fM{k}{a} - 3 \fM{j}{b} \fM{k}{b}
 \bigr)
 \biggr]
 \quad ( i \neq j \neq k \neq i),
 \label{eq:hol-Z1}
 \\
 Z^{i}_{ab} &= 
 \frac{1}{T_i+ \overline{T_i}} \biggl(
 \prod_{k=1}^3 \frac{1}{\sqrt{(U_k+\overline{U_k})}} \biggr)
 \sqrt{\frac{|I^{(i)}_{ab}|}{\prod_{j\neq i} |I^{(j)}_{ab}|}} .
 \label{eq:hol-Z2}
\end{align}
The expansion in fluxes 
is valid when $s > t_i |M^{(j)}M^{(k)}|~(i \neq j \neq k \neq i)$,
and this is similar to the case of a gauge coupling.
Then the metric $\mc{Z}^{i}_{ab}$ can be positive definite in SUSY theories
when induced D5-brane charges, 
$-M^{(j)}_a M^{(k)}_a$ and $-M^{(j)}_b M^{(k)}_b$,
are positive, even if flux corrections become large. 
This is because a sign of the product of intersection numbers, $\fI{j}{ab} \fI{k}{ab}$, 
is always positive owing to a chirality of $A_i$ multiplet.
A similar K\"{a}hler potential which depends on $S$ 
is obtained in type II theories with string scattering amplitudes \cite{Lust:2004cx,Lust:2004fi,Font:2004cx}
and is found also in Heterotic M-theory \cite{Lukas:1997fg} with an effective field theory approach.

The K\"{a}hler metrics for the other fields are systematically given by the cyclic replacement of the label of the tori and gauge groups.

\subsection{Scalar quartic term in the F-term scalar potential}

Let us check if the K\"ahler metric in the previous 
subsection is correct by showing the scalar potential.
We derive scalar quartic couplings in the F-term potential
from NDBI action and compare it with the SUGRA description.
For concreteness, we focus on 
$A_1 B_2 \ol{A_1} \ol{B_2}$ term included
in the potential.
This is related to the Yukawa coupling 
in the superpotential and hence is restricted by holomorphy.
On the other hand, there is another type of quartic terms of 
$|A_1|^4$
that is associated with D term.
The D-term scalar potential is
less constrained than that of F term
and hence we do not discuss the details in this paper for simplicity.

The leading term in flux expansion of the F-term scalar potential
which consists of multiplication of the holomorphic function and its complex conjugate one is estimated from
\begin{align*}
 2 \pi \mc{L}_{\text{4D}} \ni - V_F \ni
  2 \int d^6 y \sqrt{g_6}\, e^{4 \Phi - \varphi} g^{i\ol{i}} g^{j \ol{j}} \tr [a_i, a_j] [a_{\bar{i}}, a_{\bar{j}}].
\end{align*}
Here, $V_F$ denotes the F-term scalar potential, and we drop the covariant derivative on zero modes since we focus on a scalar quartic term.\footnote{
It is noted that $D_{z_i} a_{z_j} = \partial_{z_i} a_{z_j} + i[\hat{A}_{z_i},a_{z_j}]= 0$ for $i\neq j$ and $D_{\bar{z}_i} a_{z_i} 
= \partial_{\bar{z}_i} a_{z_i} + i[\hat{A}_{\bar{z}_i},a_{z_i}]=0$
for zero modes \cite{Cremades:2004wa,Abe:2012ya}.
Terms proportional to $D_{z_i} a_{z_i}$ and
$D_{\bar{z}_i} a_{z_j} ~(i\neq j)$ for zero modes
will contribute to 4D action as a moduli-dependent 
Fayet-Illiopoulos D-term,
which will be vanishing if the SUSY condition is preserved.
}
A factor of $e^{4 \Phi}$ originates from $\sqrt{ \tilde{g}_4}$ in the 4D effective action with the Einstein frame metric
$g_{\mu\nu} = e^{ - 2\Phi} \tilde{g}_{\mu\nu}$.
The term of $A^{\mbb{A}}_1 B^{\mbb{B}}_2 \ol{A^{\mbb{A}'}_1} \ol{B^{\mbb{B}'}_2} $ including flux corrections
arises from those proportional to $[a_1, a_2] [a_{\bar{1}}, a_{\bar{2}}]$,
\begin{align}
 V_F \ni &~ 2 \frac{ e^{3 \varphi}}{(\Vol(\mbb{T}^6))^2}\, g^{1 \ol{1}} g^{2 \ol{2}}
 \biggl[ 1 + \frac{1}{6}
 \Bigl( 
 2 \fm{1}{a} \fm{2}{a} + 2 \fm{1}{c} \fm{2}{c} + \fm{1}{a} \fm{2}{c} + \fm{1}{c} \fm{2}{a}
 \Bigr)
 \biggr]
 \nn \\
 &~ \times( \alpha^{(1)}_{ab})^2 \times (\alpha^{(2)}_{bc})^2 \times 
 A^{\mbb{A}}_1 B^{\mbb{B}}_2 \ol{A^{\mbb{A}'}_1} \ol{B^{\mbb{B}'}_2} \times
 \biggl( \int d^6 y \sqrt{g_6} \, \phi^{\mbb{A}, ab}_1 \phi^{\mbb{B}, bc}_2 
 \ol{\phi^{\mbb{A}',ab}_1}\ol{\phi^{\mbb{B}',bc}_2} \biggr)
 \\
 =& A^{\mbb{A}}_1 B^{\mbb{B}}_2 \ol{A^{\mbb{A}'}_1} \ol{B^{\mbb{B}'}_2}
 \times \frac{2 Z^3_{ca}}{\mc{Z}^3_{ca}} 
 \frac{e^{3 \varphi}}{(\Vol(\mbb{T}^6))^2} g^{1\ol{1}} g^{2 \ol{2}}
 ( \alpha^{(1)}_{ab})^2 
 (\alpha^{(2)}_{bc})^2
 \biggl( \int d^6 \sqrt{g_6} \, \phi^{\mbb{A},ab}_1 \phi^{\mbb{B},bc}_2
 \ol{\phi^{\mbb{A}',ab}_1}\ol{\phi^{\mbb{B}',bc}_2} \biggr) ,
 \label{eq:VFA1B2}
\end{align}
where Eq.~\eqref{eq:KahlermetciesZ} is used, and
$(\alpha^{(1)}_{ab})^2 \times (\alpha^{(2)}_{bc})^2$ comes from the rescaling of
$A_1^{\mbb{A}} \to \alpha^{(1)}_{ab} A^{\mbb{A}}_1$ and
$B^{\mbb{B}}_2 \to \alpha^{(2)}_{bc} B^{\mbb{B}}_2$ for the SUGRA basis.
Since $Z^3_{ca} = e^{2 \Phi - \varphi} g^{3 \ol{3}} (\alpha^{(3)}_{ca})^2$ and $\Vol(\mbb{T}^6) = e^{-2 \Phi + 2 \varphi}$,
this potential is also written as
\begin{align}
 V_F \ni  A^{\mbb{A}}_1 B^{\mbb{B}}_2 \ol{A^{\mbb{A}'}_1} \ol{B^{\mbb{B}'}_2} \times \frac{e^{K^{(0)}}}{\mc{Z}^{3}_{ca}} 
 \biggl(\sqrt{2} e^{- K^{(0)}/2} e^{3 \Phi - \varphi} \frac{\alpha^{(1)}_{ab} \alpha^{(2)}_{bc} \alpha^{(3)}_{ca}}{\sqrt{ g_{1 \ol{1}} g_{2 \ol{2}} g_{3 \ol{3}} }} \biggr)^2
 \int d^6y \sqrt{g_6} \phi^{\mbb{A}, ab}_1 \phi^{\mbb{B},bc}_2  \ol{\phi^{\mbb{A}',ab}_1}\ol{\phi^{\mbb{B}',bc}_2} .
 \label{eq:re-VF}
\end{align}
Here, 
$e^{K^{(0)}} = 1/(2^7 s t_1 t_2 t_3 u_1 u_2 u_3)$
and $g^{i \bar{i}} = 1/g_{i \bar{i}}$.

Before carrying out the integration of four zero mode functions,
we introduce a holomorphic Yukawa coupling 
$W_{\mbb{A} \mbb{B} \mbb{C}}$ with an integration of three zero mode functions, since the former integration is written as the square of the absolute value of the latter one.
As discussed in Ref.~\cite{Cremades:2004wa},
a holomorphic Yukawa coupling is expressed as
\begin{align}
 W_{\mbb{A} \mbb{B} \mbb{C}} \coloneqq&~ \sqrt{2} e^{-K_0/2} \alpha^{(1)}_{ab} \alpha^{(2)}_{bc} \alpha^{(3)}_{ca}
 \frac{e^{3 \Phi - \varphi}}{ \sqrt{ g_{1 \bar{1}} g_{2 \bar{2}} g_{3 \bar{3}} } }
 \int d^6y \sqrt{g_6} \, \phi^{\mbb{A},ab}_{1} \phi^{\mbb{B}, bc}_{2} \phi^{\mbb{C}, ca}_{3}
 \label{eq:Yukawa-int}
 \\
 =& 2 \prod_{r=1}^3 W_{A^{(r)} B^{(r)} C^{(r)}},
 \label{eq:h-Yukawa}
\end{align}
where holomorphic function of 
$W^{(r)}_{A^{(r)} B^{(r)} C^{(r)}}~(r = 1, 2,3 )$ is given by
\begin{align}
 W_{A^{(1)} B^{(1)} C^{(1)}} \coloneqq&~ \ol{\vartheta \bmat{
 \frac{B^{(1)} |I^{(1)}_{ca}| - C^{(1)} |I^{(1)}_{bc}| + m^{(1)} I^{(1)}_{bc} I^{(1)}_{ca}}{|I^{(1)}_{ab} I^{(1)}_{bc} I^{(1)}_{ca}|} \\
 0}
 (0, i \ol{U_1}| I^{(1)}_{ab} I^{(1)}_{bc} I^{(1)}_{ca}|)},
 \\
 W_{A^{(2)} B^{(2)} C^{(2)}} \coloneqq&~ \ol{
 \vartheta \bmat{\frac{C^{(2)} |I^{(2)}_{ab}| - A^{(2)} |I^{(2)}_{ca}| + m^{(2)} |I^{(2)}_{ab} I^{(2)}_{ca}|}{|I^{(2)}_{ab} I^{(2)}_{bc} I^{(2)}_{ca}|} \\
 0} (0, i \ol{U_2}|I^{(2)}_{ab} I^{(2)}_{bc} I^{(2)}_{ca}|) },
 \\
 W_{A^{(3)} B^{(3)} C^{(3)}} \coloneqq&~ \ol{ \vartheta \bmat{ \frac{A^{(3)} |I^{(3)}_{bc}| - B^{(3)} |I^{(3)}_{ab}| + m^{(3)} |I^{(3)}_{ab} I^{(3)}_{bc}|}{|I^{(3)}_{ab} I^{(3)}_{bc} I^{(3)}_{ca}|} \\
 0}
 (0, i \ol{U_3}|I^{(3)}_{ab} I^{(3)}_{bc} I^{(3)}_{ca}|)},
\end{align}
and
\begin{align}
 & A^{(1)} = B^{(1)} + C^{(1)} + m^{(1)}|I^{(1)}_{bc}|,
 \quad
 m^{(1)} = 0, 1, \ldots , I^{(1)}_{ab} -1,
 \nn \\
 & B^{(2)} = A^{(2)} + C^{(2)} + m^{(2)} |I^{(2)}_{ca}|,
 \quad
 m^{(2)} = 0, 1, \ldots, I^{(2)}_{bc} -1,
 \label{eq:yukawa-index}
 \\
 & C^{(3)} = A^{(3)} + B^{(3)} + m^{(3)} |I^{(3)}_{ab}|,
 \quad
 m^{(3)} = 0, 1,. \ldots , I^{(3)}_{ca} -1.
 \nn
\end{align}
It is noted that this coupling depends on the complex structure moduli $U_i$ via the argument of the theta function.
The coefficient in Eq.~\eqref{eq:Yukawa-int} is chosen
such that the Yukawa coupling becomes a holomorphic function 
consistent with the SUGRA formulation as noted already (see also Appendix~\ref{app:WABC}).

To evaluate the zero mode integral in the rhs of Eq.~\eqref{eq:re-VF},
we first rewrite the integral as
\begin{align}
 \nn
 & \int_{\mbb{T}^6} d^6y \sqrt{g_6} \, \phi^{\mbb{A}, ab}_1 \phi^{\mbb{B}, bc}_2 \ol{\phi^{\mbb{B}',bc}} \ol{\phi^{\mbb{A}',ab}} \\
 &=  \int_{\mbb{T}^6} d^6y \sqrt{g_6}
 \, \phi^{\mbb{A}, ab}_1 (y) \phi^{\mbb{B}, bc}_2 (y)
 \int_{\mbb{T}^6} d^6y' \sqrt{g_6}
 \, \ol{\phi^{\mbb{B}',bc}} (y') \ol{\phi^{\mbb{A}',ab}}(y')
 \times \frac{1}{\sqrt{ g_6}}\delta (y- y')
 \label{eq:4->3-1}
\end{align}
and use the following completeness relation \cite{Abe:2009dr}\footnote{
The integration on the third torus is straightforward even without the completeness relation.
We obtain the result of $|W_{A^{(3)} B^{(3)} C^{(3)}}|^2$
explicitly consistent with the SUGRA formulation 
after the integration
because both $\phi^{\mbb{A}, ab}_1$ and $\phi^{\mbb{B}, bc}_2$ have 
the (almost) antiholomorphic solution on the third torus.
}:
\begin{align}
 \sum_{n \geq 0, {\mathbb C}}\Xi_{n}^{{\mathbb C}, ca}(y) 
 \ol{\Xi_{n}^{{\mathbb C},ca}(y')} 
 = \frac{1}{\sqrt{g_6}}\delta (y- y').
 \label{eq:completeness}
\end{align}
Here, $\Xi_{n}^{\mbb{C}, ca}$ are the eigenfunctions of 
the Dirac equation with the magnetic flux of $ \fI{i}{ca} = \fM{i}{c} - \fM{i}{a}$
on each torus, and
$n$ denotes the label of the Landau level including the zero mode.
The degeneracy is given by $|I_{ca}|$.
These functions are assumed to be normalized as
\begin{align}
 \int_{\mbb{T}^6} d^6 y \sqrt{g_6} \,
 \overline{\Xi_{m}^{{\mathbb C}', ca}} 
 \Xi_{n}^{{\mathbb C},ca} = \delta_{m,n}\delta_{{\mathbb C},{\mathbb C}'} .
\end{align}
Massive modes in Landau level are orthogonal to zero modes, 
so Eq.~\eqref{eq:4->3-1} becomes
\begin{align}
 \nn
 \sum_{{\mathbb C}} \int_{\mbb{T}^6} d^6y \sqrt{g_6}
 \, \phi^{\mbb{A}, ab}_1 (y) \phi^{\mbb{B}, bc}_2 (y)
 \phi^{\mbb{C}, ca}_3 (y)
 \times  \int_{\mbb{T}^6} d^6y' \sqrt{g_6}\,
 \ol{\phi^{\mbb{A}',ab}}(y')
 \ol{\phi^{\mbb{B}',bc}} (y') 
 \ol{\phi^{\mbb{C}, ca}_3} (y') .
 \label{eq:4->3-2}
\end{align}
Thus, this is evaluated as
\begin{align}
 \int_{\mbb{T}^6} d^6y \sqrt{g_6} \, \phi^{\mbb{A}, ab}_1 \phi^{\mbb{B}, bc}_2 \ol{\phi^{\mbb{B}',bc}} \ol{\phi^{\mbb{A}',ab}}
& = \frac{1}{2^2} \frac{\sqrt{2^3 u_1 u_2 u_3}}{\Vol(\mbb{T}^6)}
 \frac{\sqrt{I_{ab} I_{bc} I_{ca}}}{\fI{1}{ab} \fI{2}{bc} \fI{3}{ca}}
 \sum_{\mbb{C}} W_{\mbb{A} \mbb{B} \mbb{C}} \ol{W_{\mbb{A}' \mbb{B}' \mbb{C}}}
 \nn \\
& = \biggl(\sqrt{2} e^{- K^{(0)}/2} e^{3 \Phi - \varphi} \frac{\alpha^{(1)}_{ab} \alpha^{(2)}_{bc} \alpha^{(3)}_{ca}}{\sqrt{ g_{1 \ol{1}} g_{2 \ol{2}} g_{3 \ol{3}} }} \biggr)^{-2}
\sum_{\mbb{C}} W_{\mbb{A} \mbb{B} \mbb{C}} \ol{W_{\mbb{A}' \mbb{B}' \mbb{C}}},
\end{align}
where flavor labels including $\mbb{A}'$ and $\mbb{B}'$ satisfy Eq.~\eqref{eq:yukawa-index} 
and a factor $1/2^2$ comes from the normalization 2 of $W_{\mbb{A} \mbb{B} \mbb{C}}$ in Eq.~\eqref{eq:h-Yukawa}.
Using this result,
Eq.~\eqref{eq:re-VF} becomes
\begin{align}
 V_F \ni
 \frac{e^{K^{(0)}}}{\mc{Z}^3_{ca}}
 \times
 A^{\mbb{A}}_1 B^{\mbb{B}}_2 \ol{A^{\mbb{A}'}_1} \ol{B^{\mbb{B}'}_2}
 \times
 \sum_{\mbb{C}} W_{\mbb{A} \mbb{B} \mbb{C}} \ol{W_{\mbb{A}' \mbb{B}' \mbb{C}}}
 .
 \label{eq:VF-A1B2}
\end{align}
Suppose that the superpotential is given by
\begin{align}
 W = \sum_{\mbb{A}, \mbb{B}, \mbb{C}}  W_{\mbb{ABC}} A^\mbb{A}_1 B^{\mbb{B}}_2 C^{\mbb{C}}_3,
 \label{eq:W}
\end{align}
where $W_{\mbb{A} \mbb{B} \mbb{C}}$ is the holomorphic Yukawa coupling defined in Eq.~\eqref{eq:Yukawa-int}.
This superpotential is discussed also in Refs.~\cite{Cremades:2004wa,Abe:2012ya}.
With this superpotential,
the above scalar potential turns out to be written based on the SUGRA formulation:
\begin{align}
V_F &\ni \frac{e^{K^{(0)}}}{\mc{Z}^3_{ca}} \sum_{\mbb{C}} \bigl( \del_{C_3^{\mbb{C}}} W)
 \ol{ \bigl( \del_{C_3^{\mbb{C}}} W \bigr)} \\
& \ni  \frac{e^{K^{(0)}}}{\mc{Z}^{3}_{ca}} \times
 A^{\mbb{A}}_1 B^{\mbb{B}}_2 
 \ol{A^{\mbb{A}'}_1} \ol{B^{\mbb{B}'}_2}
 \times
 \sum_{\mbb{C}} W_{\mbb{ABC}} \ol{W_{\mbb{A}'\mbb{B}'\mbb{C}}}   .
\end{align}
Thus, the K\"ahler metric derived from the NDBI 
action is consistent with the scalar potential based on 
the SUGRA formulation.

\section{Summary and discussions}
\label{sec:summary}

4D $\mc{N}=1$ supersymmetric effective action is 
systematically derived from the 10D NDBI action on a six-dimensional magnetized torus.
The 10D action is expanded in the series of fluxes up to ${\cal O}(F^4)$ 
with a symmetrized trace prescription.
The eigenfunctions of the Dirac equations on the torus are explicitly
written with using the Jacobi theta function and contribute 
to the 4D effective action as an integrand in the extra dimension.
We calculated the flux corrections systematically to 
the matter K\"{a}hler metrics, the gauge couplings and 
the holomorphic superpotential
via scalar quartic couplings in the F-term potential.
Our finding is a new flux correction appearing in the K\"{a}hler metrics of Eqs.~\eqref{eq:hol-Z1} and \eqref{eq:hol-Z2} in a flavor-independent way.
The new matter K\"{a}hler metric depends on the fluxes, 4D dilaton, K\"{a}hler moduli, and complex structure moduli
and will be always positive definite if an induced RR charge of the D-branes on which matters are living are positive.
A contribution of the new matter K\"{a}hler metric to the F-term scalar potential turns out to be consistent with the SUGRA formula. 
The gauge coupling in Eq.~\eqref{eq:hol-gauge} and
the holomorphic superpotential in Eq.~\eqref{eq:W} are consistent with the previous works.

Phenomenologically, matter K\"ahler metrics contribute to physical Yukawa couplings in a flavor-independent way.
If fluxes on a stack of D-branes on which quarks in the Standard Model are living
are different from those on which leptons are living,
differences in their K\"ahler metrics will be induced and could explain the mass difference between quarks and leptons.
If fluxes are common both in the quark sector and lepton one
as in the Pati-Salam like D-brane models,
such an explanation will be difficult in toroidal compactifications.
As for SUSY breaking effects to chiral matters, even if
vacuum expectation values of F components of $T_i$ and $U_i$
are much smaller than that of $S$, the flux corrections depending on $S$
in the K\"ahler metrics can generate sizable soft terms in comparison with cases without the corrections \cite{Font:2004cx}.

In this work, we consider the SUSY condition of \eqref{eq:SUSY2}.
However, if the configuration of D9-branes is supersymmetric,
this condition will be modified as
\begin{align}
 \sum_i \frac{\fM{i}{\alpha}}{\mc{A}^{(i)}} = \prod_{j=1}^3 \frac{\fM{j}{\alpha}}{\mc{A}^{(j)}},
 \quad
 \alpha = a, b, c.
\end{align}
It will be worthwhile studying the D-term potential including
the Fayet-Illiopoulos term.
Further, imposing this SUSY condition on the D9-brane action requires higher order corrections to the Lagrangian. For instance, 
${\cal O}(F^6)$ terms are required for the SUSY condition
when we focus on ${\cal O}(F^4)$ terms as in this paper. 
We could identify a part of ${\cal O}(F^6)$ then.
To include higher order interactions with derivatives 
can be important to study swampland conjectures with effective field theories.

\section*{Acknowledgments}
\noindent
%
This work is supported in part by JSPS Grant-in-Aid for Scientific Research KAKENHI 
Grant No. JP20J11901 (Y.A.)
and MEXT KAKENHI Grant No. JP19H04605 (T.K.).

\appendix

\section{Details of the calculations}
\label{app:details}

In this section,
we show the details of the calculations of the NDBI action
and use the action in Eq.~\eqref{eq:NDBI0} and the metric ansatz in Eq.~\eqref{eq:metric}.
Using these, we read the gauge couplings, matter kinetic terms, and quartic terms of the scalar potential
for the fluctuations in Eqs.~\eqref{eq:gaugefields} and \eqref{eq:chargedmatters} around the background fluxes
in Eq.~\eqref{eq:backgroundflux}.
Then, the 10D field strength in Eq.~\eqref{eq:NDBI0} is expressed as
\begin{align}
 F_{MN} =\hat{F}_{MN} + f_{MN},
\end{align}
where $\hat{F}_{MN}$ denotes the background flux 
with the background gauge field $\hat{A}_{z_i}$ and the fluctuation 
$f_{MN}$ is given by
\begin{align}
 &f_{\mu\nu} = \del_\mu a_\nu - \del_\nu a_\mu,
 \quad
 f_{\mu i}= \del_\mu a_{z_i} + i [a_\mu, a_{z_i}] + i [a_\mu, \hat{A}_{z_i}],
 \\
 & f_{ij} = \del_{z_i} a_{z_j} + i [\hat{A}_{z_i}, a_{z_j}] - \del_{z_j} a_{z_i} - i [ \hat{A}_{z_j}, a_{z_i}]
 + i [a_{z_i}, a_{z_j}],
 \\
 & f_{i \bar{j}} = \del_{z_i} a_{\bar{z}_j} + i [\hat{A}_{z_i}, a_{\bar{z}_j}] - \del_{\bar{z}_j} a_{z_i} - i [ \hat{A}_{\bar{z}_j}, a_{z_i}]
 + i [a_{z_i}, a_{\bar{z}_j}].
\end{align}
Here, $\hat{A}_{z_i}$ denotes the background gauge field, and $a_M$ denotes the fluctuation.
In addition, let us introduce the following quantity for simplicity:
\begin{align}
  \bF_j = g^{j \ol{j}} \hat{F}_{j \ol{j}} = i \frac{2 u_j}{\ls^2 \mc{A}^{(j)}} \frac{\pi}{u_j} M^{(i)} = \frac{i}{2\pi \ap} m^{(j)};
  \quad
  m^{(j)} \coloneqq \frac{M^{(j)}}{\mc{A}^{(j)}},
\end{align}
where $ 
j =1,2,3
$, the summation with respect to 
$j$
is not taken,
and 
$M^{(j)}$
is given by Eq.~\eqref{eq:backgroundflux}.
The SUSY condition \eqref{eq:SUSY3-1} is rewritten 
as the condition of 
$\bF_j$
as
\begin{align}
 \bF_1 + \bF_2 + \bF_3 = 0.
 \label{eq:SUSY4}
\end{align}

In the following parts,
we focus just on $f_{\mu \nu}$,
$ \partial_\mu a_{z_i} \in f_{\mu i}$,
$i [a_{z_i}, a_{z_j}] \in f_{ij}$
and $i [a_{z_i}, a_{\bar{z}_j}] \in f_{i \bar{j}}$ 
to calculate the effective action.
Derivative terms of $D_{z_i} a_{z_j} = \partial_{z_i} a_{z_j} + i[\hat{A}_{z_i},a_{z_j}]~(i\neq j)$ 
and $D_{\bar{z}_i} a_{z_i} 
= \partial_{\bar{z}_i} a_{z_i} + i[\hat{A}_{\bar{z}_i},a_{z_i}]$
are vanishing for zero modes \cite{Cremades:2004wa,Abe:2012ya}.
Terms proportional to $D_{z_i} a_{z_i}$ and
$D_{\bar{z}_i} a_{z_j} ~(i\neq j)$ for zero modes
will contribute to 4D action as a moduli-dependent 
Fayet-Illiopoulos D-term,
which will be vanishing if the SUSY condition is preserved.

\subsection{Gauge couplings}

The gauge coupling is read from the coefficient of the gauge kinetic term.
Due to the index structure of Eq.~\eqref{eq:NDBI0},
only its third and forth terms in $\mc{O}(F^4)$ contribute to 
the gauge kinetic terms, and 
then the expansion of the NDBI action is calculated as
\begin{align}
 2 \pi \mc{L}_\NDBI \ni& -\int d^6y\sqrt{g_6}\, \frac{e^{4\Phi -\varphi}}{4}
 e^{-4 \Phi} \biggl[ \tr f_{\mu\nu} f^{\mu\nu} 
 \nn \\
 &~~~~~~~~~~
 +2 \times \frac{(2 \pi \ap)^2}{8} \frac{1}{3} \tr \Bigl( 2 \hat{F}_{j \bar{k}} \hat{F}^{j \bar{k}} f_{\mu\nu}  f^{\mu\nu} + \hat{F}_{j \bar{k}} f_{\mu\nu} \hat{F}^{j \bar{k}} f^{\mu\nu} \Bigr) \biggr]
 \\
 =&
 - \int d^6 y\sqrt{g_6} \, \frac{e^{- \varphi}}{4}
 \biggl[ \tr f_{\mu\nu} f^{\mu\nu} 
 - \frac{(2\pi \ap)^2}{6} \sum_{k}
 \tr \Bigl(  2 \bF_k \bF_k f_{\mu\nu} f^{\mu\nu} + \bF_k f_{\mu\nu} \bF_k f^{\mu\nu} \Bigr) \biggr],
\end{align}
where we used the fact that the background flux is diagonal,
$\hat{F}_{j \bar{k}} = \hat{F}_{k \bar{k}} \delta_{kj}$.
Since these fluxes are assumed to be Abelian,
the Lagrangian reduces to
\begin{align}
  2 \pi \mc{L}_\NDBI &\ni 
  - \int d^6y \sqrt{g_6} \frac{e^{-\varphi}}{4}
  \tr \biggl[ \Bigl(
  1 - \frac{(2 \pi \ap)^2}{2} \sum_k \bF_k \bF_k \Bigr) f_{\mu\nu} f^{\mu\nu} \biggr] \\
  &
  = - \frac{s}{4}
  \tr \biggl[ \Bigl(
  1 + \frac{1}{2} \sum_k (m^{(k)})^2  \Bigr) f_{\mu\nu} f^{\mu\nu} \biggr],
\end{align}
where we used $\int d^6y \sqrt{g_6} e^{-\varphi} = s$.

\subsection{Kinetic terms}

The scalar kinetic terms come from those proportional to 
$f_{\mu i} f_{\nu \bar{i}} g^{\mu\nu} g^{i \bar{i}}$.
Such terms including flux corrections are given by
\begin{align}
 2 \pi \mc{L}_\NDBI \ni &
 -\int d^6 y\sqrt{g_6}\, \frac{e^{2 \Phi - \varphi}}{4}
 \biggl[ 4 \sum_i \tr f_{\mu i} f_{\nu \bar{i}} g^{\mu\nu} g^{i \bar{i}}
 \nn \\
 &~- \frac{(2\pi \ap)^2}{3} Z_{K1} - \frac{(2\pi \ap)^2}{6} Z_{K_2}
 + \frac{(2\pi \ap)^2}{12} Z_{K3} + \frac{(2 \pi \ap)^2}{24} Z_{K_4}
 \biggr],
 \label{eq:kinetiterms-app}
\end{align}
where 
\begin{align}
 Z_{K1} = & - \sum_{i} g^{i \bar{i}} g^{\mu \nu} \tr \bigl[
 2 \bF_i \bF_i
 ( f_{\mu i} f_{\nu \bar{i}} + f_{\mu \bar{i}} f_{\nu i})
 + 4 \bF_i f_{\mu i} \bF_i f_{\nu \bar{i}}
 \bigr] ,
 \\
 Z_{K2} = &-4 \sum_i g^{i \bar{i}} g^{\mu \nu} \tr \bigl[ 
 \bF_i \bF_i ( f_{\mu i} f_{\nu \bar{i}} + f_{\mu \bar{i}} f_{\nu i} )
 \bigr],
 \\
 Z_{K3} = &-8 \sum_{i} \sum_{k} g^{i \bar{i}} g^{\mu\nu} \tr \bigl[
 \bF_k \bF_k
 ( f_{\mu i} f_{\nu \bar{i}} + f_{\mu \bar{i}} f_{\nu i})
 \bigr] ,
 \\
 Z_{K4} = & -16 \sum_{i} \sum_{k} g^{i \bar{i}} g^{\mu \nu} \tr \bigl[
 \bF_k f_{\mu i} \bF_k f_{\nu \bar{i}}
 \bigr].
\end{align}

\subsubsection{K\"{a}hler metric of charged matters}

With the background and the fluctuations substituted into 
the above equations, it turns out that
the K\"{a}hler metric of chiral matter $A_i$ is given by 
\begin{align}
 2 \pi \mc{L}_\NDBI \ni& - \frac{2 u_i}{t_i \Vol(\mbb{T}^6)} \biggl[ 1 - \frac{t_i}{6s}
 \bigl( 2 \fM{j}{a} \fM{k}{a} + 2 \fM{j}{b} \fM{k}{b} + \fM{j}{a} \fM{k}{b} + \fM{j}{b} \fM{k}{a} \bigr)
 \biggr]
 |\del_\mu A_i|^2
\end{align}
with $ i \neq j \neq k \neq i$ and $i = 1$ for a fixed choice of intersection number in this paper.
The K\"{a}hler metrics for the other fields are systematically given by the cyclic replacement of the label of the tori and gauge groups.

\subsubsection{K\"{a}hler metric of open string moduli}

A diagonal part of gauge fluctuation $a_{i}$ 
is open string modulus
$a_i^{b} \coloneqq a_i^{bb}$.
Its K\"{a}hler metric can be read from Eq.~\eqref{eq:kinetiterms-app},
\begin{align}
  2 \pi \mc{L}_\NDBI & \ni - \frac{2}{(2t_i) (2u_i)}\biggl[ 1 - \frac{t_i}{s} \fM{j}{b} \fM{k}{b} \biggr]
  |\del_\mu a^{b}_i|^2,
  \\
  {\cal Z}_{bb}^i &= 
  \frac{2}{(T_i+\bar{T}_i) (U_i +\bar{U}_i)}
  \biggl[ 1 - \frac{(T_i+\bar{T}_i)}{(S+\bar{S})}
  \fM{j}{b} \fM{k}{b} \biggr] .
\end{align}
It is noted that a flux correction in this result is
obtained also by replacing $M_a$ with $M_b$.
This matches the result discussed in Refs.~\cite{Kors:2003wf,Lust:2004cx,Lust:2004fi,Font:2004cx}.
The positivity condition on the kinetic term of the open string modulus is same as that of the gauge coupling.
%

\subsection{Quartic terms}

Scalar quartic terms originate from
those including $g^{i \bar{i}} g^{j \bar{j}}  \bigl(
f_{ij} f_{\bar{i} \bar{j}} +  f_{i\bar{j}} f_{\bar{i} j} \bigr)$,
where $f_{ij} \coloneqq i [a_i, a_j]$ and 
$f_{i \ol{j}} \coloneqq i [a_i , a_{\bar{j}}]$.
We can read such terms from NDBI action,
\begin{align}
 2 \pi \mc{L}_\NDBI \ni & - \int d^6y \sqrt{g_6}\, \frac{e^{4 \Phi - \varphi}}{4}
 \biggl[ \sum_{i, j} 2 g^{i \bar{i}} g^{j \bar{j}}  \bigl(
 \tr f_{ij} f_{\bar{i} \bar{j}} + \tr f_{i\bar{j}} f_{\bar{i} j} \bigr)
 \nn \\
 &~ - \frac{(2\pi \ap)^2}{3} K_1 - \frac{(2\pi \ap)^2}{6} K_2 + \frac{(2\pi \ap)^2}{12} K_3 + \frac{(2\pi \ap)^2}{24} K_4 
 \biggr],
 \label{eq:quartic-app}
\end{align}
and $K_{p=1,2,3,4}$ are decomposed to two parts: 
one is $K_{p,F}$ containing $f_{ij}f_{\bar{i} \bar{j}}$
and the other is $K_{p,D}$ containing $f_{i\bar{j}} f_{\bar{i} j}$.
Explicitly they are given by
\begin{align}
 K_{1,F}
 =& \sum_{i, j} g^{i \ol{i}} g^{j \bar{j}} \tr \Bigl\{
  2 \bF_i \bF_i ( f_{ij} f_{\ol{ji}} + f_{\ol{ij}} f_{ji} )
 + 4 \bF_i f_{ij} \bF_i f_{\ol{ji}}
 \nn \\
 &~ - ( \bF_i \bF_j + \bF_j \bF_i ) ( f_{ij} f_{\ol{ji}} + f_{\ol{ij}} f_{ji} )
 \Bigr\}
 ,
 \\
 K_{1,D}
 =& \sum_{i,j} g^{i \ol{i}} g^{j \ol{j}} \tr \Bigl\{
 2 \bF_i \bF_i ( f_{j \ol{i}} f_{i \ol{j}} + f_{\ol{j} i} f_{\ol{i} j} ) + 4 \bF_i f_{i \ol{j}} \bF_i f_{j \ol{i}}
 \nn \\
 &~ + (\bF_i \bF_j + \bF_j \bF_i) ( f_{i \ol{j}} f_{j \ol{i}} + f_{\ol{i} j} f_{\ol{j} i} )
 \Bigr\}
 ,
 \\
 K_{2,F}
 =& \sum_{i,j} g^{i \ol{i}} g^{j \ol{j}} \tr \Bigl\{
 4 \bF_i \bF_i ( f_{ij} f_{\ol{ji}} + f_{\ol{ij}} f_{ji}) - 4 \bF_i f_{ij} \bF_j f_{\ol{ji}}
 \Bigr\}
 ,
 \\
 K_{2,D}
 =& \sum_{i,j} g^{i \ol{i}} g^{j \ol{j}} \tr \Bigl\{
 4 \bF_i \bF_i ( f_{i \ol{j}} f_{j \ol{i}} + f_{\ol{i} j} f_{\ol{j} i} )
 +2 ( \bF_i f_{i \ol{j}} \bF_j f_{j \ol{i}} + \bF_i f_{\ol{i} j} \bF_j f_{\ol{j} i} )
 \Bigr\}
 ,
 \\
 K_{3,F}
 =& \sum_{i,j} g^{i \ol{i}} g^{j \ol{j}} \tr \Bigl\{
 4 \bigl( \sum_k \bF_k \bF_k \bigr) ( f_{ij} f_{\ol{ji}} + f_{\ol{ij}} f_{ji} )
 \Bigr\}
 ,
 \\
 K_{3,D}
 =& \sum_{i,j} g^{i \ol{i}} g^{j \ol{j}} \tr \Bigl\{
 4 \bigl( \sum_k \bF_k \bF_k \bigr) ( f_{i \ol{j}} f_{j \ol{i}} + f_{\ol{i} j} f_{\ol{j} i})
 \nn \\
 &~ + 4 \bigl(
 \bF_i f_{i \ol{i}} \bF_j f_{j \ol{j}} + \bF_i \bF_j f_{j \ol{j}} f_{i \ol{i}} + f_{i \ol{i}} \bF_i f_{j \ol{j}} \bF_j + \bF_i f_{i \ol{i}} f_{j \ol{j}} \bF_j
 \bigr)
 \Bigr\}
 \\
 K_{4,F}
 =& \sum_{i,j} 8 g^{i \ol{i}} g^{j \ol{j}} \tr \Bigl\{
 \sum_k \bF_k f_{ij} \bF_k f_{\ol{ji}}
 \Bigr\}
 ,
 \\
 K_{4,D}
 =& \sum_{i,j} g^{i\ol{i}} g^{j \ol{j}} \tr \Bigr\{
 8 \sum_k \bF_k f_{i \ol{j}} \bF_k f_{j \ol{i}} + 4\bigl( \bF_i \bF_j f_{i \ol{i}} f_{j \ol{j}}
 \nn \\
 &~
 + \bF_i f_{j \ol{j}} f_{i \ol{i}} \bF_j + f_{i \ol{i}} \bF_j \bF_i f_{j \ol{j}} + f_{i \ol{i}} f_{j \ol{j}} \bF_i \bF_j \bigr)
 \Bigr\}
 .
\end{align}
If we want to get the specific quartic coupling such as $A^{\mbb{A}}_1 B^{\mbb{B}}_2 \ol{B^{\mbb{B}'}_2} \ol{A^{\mbb{A}'}_1}$,
one needs to choose a term with a fixed index like $f_{12}f_{\bar{1}\bar{2}}$.

\subsubsection{F-term potential}

The terms containing $f_{ij}~(i \neq j)$ contribute to the scalar F-term potential.
When the background fluxes satisfy the SUSY condition of Eq.~\eqref{eq:SUSY3-1} or \eqref{eq:SUSY4},
we can show that 
$K_{p,D}~(p=1,2,3,4)$ does not include $f_{i \ol{j}}~(i\neq j)$
and hence does not contribute to the F-term scalar potential.
In the leading contribution in the flux expansion, however,
$\tr [a_i, a_{\ol{j}}] [ a_j, a_{\ol{i}}]$ is shown to have
$\tr [a_i, a_j] [a_{\ol{i}}, a_{\ol{j}}] $ contributing to the F-term potential through the Jacobi identity as
\cite{Witten:1985xb}
\begin{align}
 \tr [ a_i, a_{\ol{j}} ] [a_j ,a_{\ol{i}} ] + \tr [a_i , a_{\ol{i}} ][ a_{j} , a_{\ol{j}} ]
 + \tr [a_i, a_j] [a_{\ol{i}}, a_{\ol{j}}] =0.
 \label{eq:Jacobiforf}
\end{align}
Then, the F-term potential from NDBI action is given by
\begin{align}
 2 \pi \mc{L}_\NDBI \ni &~ - \frac{1}{4} \int d^6y\sqrt{g_6} \, e^{4 \Phi - \varphi}
 \sum_{i < j} g^{i \ol{i}} g^{j \ol{j}} \tr \biggl\{
 8 f_{ij} f_{\ol{i} \ol{j}} 
 - \frac{4(2 \ap)^2}{3} \Bigl[ \bF_i f_{ij} \bF_j f_{\ol{i} \ol{j}} + \bF_j f_{ij} \bF_i f_{\ol{i} \ol{j}}
 \nn \\
 &~ + ( \bF_i \bF_j + \bF_j \bF_i ) (f_{ij} f_{\ol{i} \ol{j}} + f_{\ol{i} \ol{j}} f_{ij} ) \Bigr]
 \biggr\}.
\end{align}
Thus, we can get Eq.~\eqref{eq:VFA1B2} 
by substituting the flux background and fluctuations into this Lagrangian.

\section{Comments on the Yukawa type superpotential}
\label{app:WABC}

Here, we show that a factor 2 in Eq.~\eqref{eq:h-Yukawa}
is consistent with the SUGRA formulation.
Let us consider the following Yukawa type superpotential with introduction of a coefficient $w$:
\begin{align}
 W_w= w {\bm{\vartheta}_{\mbb{A} \mbb{B} \mbb{C}}} A_1^{\mbb{A}} B_2^{\mbb{B}} C_3^{\mbb{C}}
 = \frac{w}{\sqrt{2}} \frac{ e^{-K_0/2} \alpha^{(1)}_{ab} \alpha^{(2)}_{bc} \alpha^{(3)}_{ca} e^{3 \Phi - \varphi}}{\sqrt{ g_{1\ol{1}} g_{2\ol{2}} g_{3\ol{3}}}} \int d^6 \sqrt{g_6} \phi^{\mbb{A}, ab}_{1} \phi^{\mbb{B},bc}_2 \phi^{\mbb{C},ca}_3 A^{\mbb{A}} B^{\mbb{B}} C^{\mbb{C}},
\end{align}
where 
\begin{align}
 {\bm{\vartheta}_{\mbb{A} \mbb{B} \mbb{C}}} \coloneqq \prod_{r=1}^3 W_{A^{(r)} B^{(r)} C^{(r)}}.
\end{align}
With this superpotential,
the scalar potential of $|A^{\mbb{A}}_1 B^{\mbb{B}}_2|^2$ 
derived from the NDBI action
is expressed as
\begin{align}
& A^{\mbb{A}}_1 B^{\mbb{B}}_2 \ol{A^{\mbb{A}'}_1} \ol{B^{\mbb{B}'}_2}
 \times
 \frac{2}{\mc{Z}^3_{ca}} e^{K_0}
 \biggl(\frac{ e^{-K_0/2} \alpha^{(1)}_{ab} \alpha^{(2)}_{bc} \alpha^{(3)}_{ca} e^{3 \Phi - \varphi}}{\sqrt{ g_{1\ol{1}} g_{2\ol{2}} g_{3\ol{3}}}} \biggr)^2 \int d^6y \sqrt{g_6} 
 \phi^{\mbb{A}, ab}_1 \phi^{\mbb{B}, bc}_2 \ol{\phi^{\mbb{B}',bc}_2} \ol{\phi^{\mbb{A}',ab}_1}
 \\
 \in & \biggl( \frac{2}{w} \biggr)^2 \frac{1}{\mc{Z}^3_{ca}} e^{K_0} |\del_{C_3} W_w|^2
 ,
\end{align}
which implies that $w =2$ makes this SUGRA potential be equal to Eq.~\eqref{eq:re-VF} derived from NDBI action.

\newcommand{\arxivfont}{\rmfamily}
\bibliographystyle{yautphys}
\bibliography{ref}

\end{document}